\def \abc#1#2#3#4 {\reference#1, {\sl#2}, {\bf#3}, #4}
\def \blank {\lower 5pt\hbox to 0.75in{\hrulefill}}

\def \cm{~\rm{cm}}

\def \K{~\rm{K}}
\def \g{~\rm{g}}

\def \AU{~\rm{AU}}

\def \yrs{~\rm{yrs}}
\def \yr{~\rm{yr}}

\def\astrobj#1{#1}
\documentclass[12pt,preprint]{aastex}
\usepackage{natbib}

\shorttitle{Opacity and Over-expansion{Soker}}

\begin{document}

\title{PHOTOSPHERIC OPACITY AND OVER-EXPANDED ENVELOPES OF
 ASYMPTOTIC GIANT BRANCH STARS}
\author{Noam Soker \altaffilmark{}}
\affil{Department of Physics, Technion$-$Israel Institute of Technology,
Haifa 32000 Israel;
soker@physics.technion.ac.il}

\begin{abstract}

I suggest that the behavior of the photospheric opacity in
oxygen-rich (similar to solar abundance) upper asymptotic giant
branch stars may cause these stars to substantially expand for a
few thousand years. I term this process over-expansion. This may
occur when the photospheric (effective) temperature drops to $T_p
\sim 3000 \K$, and because the opacity sharply increases as
temperature further decreases down to $T_p \sim 2000 \K$.
The much higher opacity implies a much lower photospheric density,
which stabilizes the envelope structure. As mass loss proceeds, the star
eventually contracts to become a post-asymptotic giant branch
star.
Some possible outcomes of the over-expanded phase are discussed:
(1) The over-expanded phase may be connected to the
formation of semi-periodic concentric arcs (rings; shells); (2)
The over-expanded phase may be related to the positive correlation
between the mass loss rate and the transition to axisymmetric mass
loss geometry; and (3) An over-expanded asymptotic giant branch star,
which doubles its radius, is somewhat more
likely to swallow a low mass companion.
\end{abstract}



\section{INTRODUCTION}

The transition from spherically symmetric mass loss to
axisymmetric mass loss, occurring in many stars evolving close to
the tip of the asymptotic giant branch (AGB) on the
Hertzsprung-Russell diagram, is poorly understood. This transition
is manifested most clearly in the structures of many elliptical
planetary nebulae (PNs). In these PNs the inner region, composed
of a shell and a rim which were formed from mass loss episodes
near the tip of the AGB and during the post-AGB phases (e.g.,
Frank, Balick, \& Riley 1990), is axisymmetric rather than
spherical, while the halo, which is composed of mass blown
somewhat earlier on the AGB, has a large scale spherical structure
(Balick et al. 1992; Corradi et al. 2003), although it may still
possess many small scale filaments, blobs, dents, etc. Examples of
such PNs are \astrobj{NGC 6543}, \astrobj{NGC 6826} (e.g., Balick 1987;
Corradi et al. 2003), and \astrobj{NGC 6891} (Guerrero et al. 2000).
In a recent paper
Corradi et al. (2003) conduct a thorough study of such PNs, giving
more examples, e.g., NGC 3918, and analyzing some properties of
the PNs shells and halos. That study sharpens the old question of
what is (are) the mechanism(s) responsible for the transition from
spherical to axisymmetric mass loss geometry near the tip of the AGB.
The large sample of PNs with spherical, or almost spherical,
halos shows that the transition to axisymmetric mass loss is
accompanied by a much higher mass loss rate (Corradi et al. 2003),
i.e., a final intensive wind (FIW; or superwind).
>From the density structure of NGC 6826 which
was deconvolved by Plait and Soker (1990), I find that the mass
loss rate into the inner elliptical shell was $\sim 10 $ times
that into the spherical halo.
This estimate is crude,
however, because evolution significantly changes the density
profile in the envelope (e.g., Sch\"onberner \& Steffen 2002).
The connection between a very high mass loss rate and the
transition to axisymmetric mass loss is evident also from the
relatively faint and smooth structure of spherical PNs (Soker
2002b).

In the past I proposed two mechanisms for the rapid transition to
axisymmetric and higher mass loss rate wind.
In the first mechanism (Soker 1995) a companion, which can be a massive
planet, a brown dwarf, or a low mass main sequence star, spins-up the
envelope, leading both to axisymmetric mass loss geometry and to a
higher mass loss rate.
The spun-up envelope blows axisymmetric wind. The interaction
of the companion with the AGB core may lead also to the formation of
jets (Soker \& Livio 1994; Soker 1996; Reyes-Ruiz \& Lopez 1999),
which forms the FLIERs (fast low ionization emission regions;
also termed {\it ansae}).
The strong dependence of tidal interaction on the AGB radius to
orbital separation ratio implies that most of the envelope spin-up
occurs in a very short time (Soker 1995).
The second mechanism (Soker 1998; Soker \& Harpaz 1999)
is based on enhanced dust formation above
cool spots on the AGB surface, most likely magnetic cool spots.
The axisymmetric mass loss is caused by an optically thick wind.
When the mass loss rate increases to the stage that the dust shields the
region above it from most of the stellar radiation, further dust formation
occurs in regions above the spots, leading to a much higher mass loss rate
above these spots (Soker 2000b).
If the cool spots are concentrated in the equatorial plane, an
axisymmetric mass loss will commence.
Here the transition to axisymmetric mass loss is caused by
an optically thick wind, rather than the onset of a binary interaction.
However, a binary companion is still required to spin-up the envelope to
define the symmetry axis and to form jets, but it may spin-up the envelope
at earlier stages, during which the mass loss rate is too low for
transition to an axisymmetric mass loss.

Motivated by the common occurrence of spherical, or almost spherical,
faint halos around inner axisymmetric regions of elliptical PNs
(Corradi et al.\ 2003; Balick et al. 1992),  I explore another
possible mechanism that may influence the rapid change in the mass
loss geometry and rate of stars evolving near the AGB tip.
This mechanism is based on the steep change in the behavior of the
opacity of solar composition gas at a temperature of $\sim 2900 \K$
(Sec. 2), and it may coexist with, and increase the efficiency of,
the other two mechanisms mentioned above.
The mechanism may operate in AGB stars having an oxygen to carbon
abundance ratio of O/C$~\gtrsim 1.1$.
It is not important in AGB stars with a carbon to oxygen
abundance ratio of C/O$~\gtrsim 0.95$, whose opacity behaves differently
(Marigo 2002, 2003).
Many of the relevant PNs discussed here have indeed O/C$~\gtrsim 1.5$,
e.g., \astrobj{NGC 6826}, \astrobj{NGC 6543} (Quigley \& Bruhweiler 1995;
Guerrero \& Manchado 1999), \astrobj{NGC 7009}, and
\astrobj{NGC 3242} (Perinotto \& Benvenuti 1981).
In Section 3 I discuss possible implications of the proposed mechanism.
A short summary is in Section 4.

\section{OPACITY AND THE PHOTOSPHERIC DENSITY}
\subsection{General behavior on the AGB}

The photospheric density is given by (Kippenhahn \& Weigert 1990)
\begin{equation}
\rho_p = \frac{2}{3} \frac {\mu m_H}{k_B} \frac {G M_\ast}{R^2 \kappa T_p},
\end{equation}
where $M_\ast$ and $L$ are the stellar mass and luminosity,
$T_p$ is the photospheric (effective) temperature,
$k_B$ is the Boltzmann constant, and $\kappa$ is the opacity.
To incorporate both low and high temperatures, I scale the mean
mass per particle by  $\mu m_H =2 \times 10^{-24} \g$, and
use it throughout the paper unless otherwise mentioned explicitly.
Substituting typical values for upper AGB stars in the last equation,
and expressing the radius in terms of the luminosity and temperature,
give
\begin{equation}
\rho_p= 6 \times 10^{-10}  
\left( \frac{\kappa}{10^{-3} \cm^2 \g^{-1}} \right)^{-1}
\left( \frac{T_p}{3000 \K} \right)^{3}
\left( \frac{L}{10^4 L_\odot} \right)^{-1}
\left( \frac{M_\ast}{1 M_\odot} \right)
\left( \frac{\mu m_H}{2 \times 10^{-24}} \right) \g \cm^{-3}.
\end{equation}
The average density in the envelope is meaningful as long as
the envelope mass is $M_{\rm env} \gtrsim 0.01 M_\odot$.
At lower envelope mass most of the mass is concentrated
near the core.
The average envelope density is
\begin{equation}
\rho_a= \frac{M_{\rm env}}{4 \pi R^3/3} = 8 \times 10^{-9}  
\left( \frac{T_p}{3000 \K} \right)^{6}
\left( \frac{L}{10^4 L_\odot} \right)^{-3/2}
\left( \frac{M_{\rm env}}{0.3 M_\odot} \right)  \g \cm^{-3}.
\end{equation}
The ratio of photospheric to average density is then
(I omit the dependence on $\mu m_H$ for the rest of the paper).
\begin{equation}
\frac{\rho_p}{\rho_a} \simeq 0.08
\left( \frac{\kappa}{10^{-3} \cm^2 \g^{-1}} \right)^{-1}
\left( \frac{T_p}{3000 \K} \right)^{-3}
\left( \frac{L}{10^4 L_\odot} \right)^{1/2}
\left( \frac{M_\ast}{1 M_\odot} \right)
\left( \frac{M_{\rm env}}{0.3 M_\odot} \right)^{-1}.
\end{equation}

To derive the dependence of opacity on temperature for
solar composition PNs, or more accurately for O/C$~\gtrsim 1.5$,
I use the results of  Alexander \& Ferguson (1994) and Ferguson
et al.\ (2005).
The opacity depends on the density, and the photospheric
density is given by equation (1), implying an implicit dependence,
which depends also on the stellar luminosity and mass.
In the range of the effective temperatures
$3000 \lesssim T \lesssim 3600 \K$ and for photospheric densities
appropriate for the upper AGB, Soker \& Harpaz (1999)
\footnote{Note that the density scale in Figs.\ 1-5 of
Soker \& Harpaz (1999) is too low by a factor of 10;
the correct scale is displayed in their Fig. 6.}
approximate the dependence of opacity on photospheric
temperature by  $\kappa \sim T^4$.
The opacity sharply changes its behavior at $T \sim 2800 \K$
(Fig. 2 of Alexander \& Ferguson 1994; Fig. 9 of Ferguson et al.\ 2005).
I solve equation (2) using the opacity table of Alexander \& Ferguson
(1994), simultaneously for the density and opacity at two points:
at $T_p=2800 \K$ I find  $\rho= 8 \times 10^{-10} \g \cm^{-3}$
and $\kappa= 6.5 \times 10^{-4} \cm^2 \g^{-1}$, while at
$T_p=2000 \K$ I find $\rho= 2 \times 10^{-11} \g \cm^{-3}$ and
$\kappa= 0.01 \cm^2 \g^{-1}$.
In this temperature range, therefore, on average $\kappa \sim T^{-8}$.
The sharp rise in opacity as temperature drops results
from molecule formation, mainly H$_2$O and TiO
(Alexander \& Ferguson 1994).
At relevant lower temperatures the opacity does not depend strongly
on the density, and it is given by $\kappa \sim T^4$.
Overall, a good approximation for the variation of the
photospheric opacity in upper AGB stars is given by
\begin{eqnarray}
\qquad  0.01 \left(\frac {T_p}{1800 \K} \right)^4, \qquad
               {\rm for} \qquad 1600 \lesssim T \lesssim 1900
\cr
\kappa (T) \simeq
 0.01 \left(\frac {T_p}{2000 \K} \right)^{-8}, \qquad
               {\rm for} \qquad 1900 \lesssim T \lesssim 2900
\cr
\qquad  5 \times 10^{-4} \left(\frac {T_p}{3000 \K} \right)^4, \qquad
               {\rm for} \qquad 2900 \lesssim T \lesssim 3600 .
\end{eqnarray}
It should be noted that the approximation above holds for upper AGB
stars, and it includes already the implicit dependence on the density.
(For the explicit dependence on density and temperature in the middle
temperature range above, for example, $\kappa (\rho, T) \sim \rho T^{-16}$,
for the relevant densities and temperatures.)

Using equation (5) in the intermediate temperature range,
and keeping parameters as in equation (2), yields for
the photospheric density, and its ratio to the average density,
\begin{equation}
\rho_p \simeq 2 \times 10^{-11}  
\left( \frac{T_p}{2000 \K} \right)^{11}
\left( \frac{L}{10^4 L_\odot} \right)^{-1}
\left( \frac{M_\ast}{1 M_\odot} \right) ,
\qquad {\rm for} \qquad 1900 \lesssim T \lesssim 2900 ,
\end{equation}
and
\begin{equation}
\frac{\rho_p}{\rho_a} \simeq 0.03 
\left( \frac{T_p}{2000 \K} \right)^{5}
\left( \frac{L}{10^4 L_\odot} \right)^{1/2}
\left( \frac{M_\ast}{1 M_\odot} \right)
\left( \frac{M_{\rm env}}{0.3 M_\odot} \right)^{-1} ,
\qquad {\rm for} \qquad 1900 \lesssim T \lesssim 2900 ,
\end{equation}
respectively.

\subsection{Evolution at low temperatures: over-expansion}
To possess a stable atmosphere, the photospheric density must be much
below average density (although a large density inversion might be
presence in the outer parts of the envelope; Soker \& Harpaz 2002).
As the star evolves along the AGB its
luminosity increases and its envelope mass decreases.
By equation (4), to maintain a ratio of $\rho_p/\rho_a \ll 1$,
the opacity  and/or the temperature should increase.
Both opacity and temperature increase
as the star shrinks during the post-AGB, where $T_p>3000 \K$.
The behavior of the opacity in the temperature range
$2000 \lesssim T_p \lesssim 2900 \K$, as given in equation (5),
implies, as is evident from equation (7), that in AGB stars with
O/C$~\gtrsim 1.1$ the densities ratio $\rho_p/\rho_a$
can be also kept very small if the star expands
in this temperature range.
In carbon stars, i.e., C/O$~>1$, the opacity, because of
CN molecules, increases already at $T_p \sim 4000 \K$.
As noted here, higher opacity implies larger radius.
Marigo (2002; 2003) discusses the need to include the enhanced
opacity in carbon stars to explain their average much larger
radii and lower temperatures than those of oxygen-rich stars.
As I showed above, the large increase in radius of oxygen-rich
stars will occur only at $T_p \lesssim 2900 \K$;
below I argue that this expansion has a different nature
from the gradual expansion of carbon-rich AGB stars.
Many PNs were formed from oxygen-rich stars
(e.g, Perinotto \& Benvenuti 1981), including PNs with semi-periodic
concentric rings, e.g., NGC 6543 (Guerrero \& Manchado 1999),
and PNs with a spherical halo, e.g., NGC 6826
(Quigley \& Bruhweiler 1995; Guerrero \& Manchado 1999).
These PNs are the subject of the present study.

The study of the exact evolution of stars in this temperature range
requires numerical simulations with accurate opacities and
the inclusion of dynamic effects.
In the present exploratory paper I proposed the following.
An AGB star reaching the evolutionary point where the photospheric
temperature is $T_p \sim 3000 \K$ and its envelope density is
very low, such that the density ratio becomes
$\rho_p/\rho_a \gtrsim 0.1$, might reduce this ratio and stabilize
its atmosphere either by contracting, or by expanding.
I term this expansion {\it over-expansion}.
The over-expansion must be triggered by strong disturbances,
and it last for a relatively short time.
Eventually the star must shrink as it evolves toward the post-AGB
phase. However, for some evolutionary time the star may
be over-expanded.
Most likely, as the star starts its over-expanding
phase it will expand all the way to $T_p \simeq 2000-2400 \K$
in a short time.
As is evident from equation (6), this is because the photospheric
density drops sharply as the star over-expands.
The shallow density profile below the photosphere when the
envelope mass is low (Soker \& Harpaz 1999), and the
fast decrease in photospheric density during the over-expansion,
$\rho_p \sim R^{-5.5}$ (see below), mean that the total mass below the
photosphere decreases.
Therefore, while some envelope mass is moving outward, some
mass moves inward.
Not much energy is needed, therefore, to raise material
to larger radii, and it may even be that with the drop
in thermal energy as the temperature drops, with recombination,
with molecule formation, and with the motion inward mentioned above,
the envelope releases energy.

To examine the change in the thermal plus gravitational energy
of the over-expanded envelope, and whether indeed this change is small,
and sometimes even energy is released, a full numerical code is required,
including dynamical effects and opacity at low temperatures.
I now present a simple calculation which suggests that energy can be released by
the envelope, hence very high mass loss rate can then remove the outer
layers, causing the envelope to shrink back to its normal state.

Based on the density profiles of AGB stars with low envelope mass
(Soker \& Harpaz 1999), I take the density profile from the photosphere
at radius $R_p$ down to radius $r \sim 0.3-0.5 R_p$ to be
\begin{equation}
{\rho(r)} = \rho_p \left( \frac {r}{R_p} \right)^{-d},
\end{equation}
where $d$ is a constant.
For low mass envelopes of stars about to leave the AGB, i.e.,
having envelope mass of $M \sim 0.1-0.2 M_\odot$,
Soker \& Harpaz (1999) density profiles can be crudely fitted with
$d \sim 4-5$.
Let us consider a star experiencing over-expansion in the
temperature range where equation (6) applies.
Let the photospheric density at stellar radius $R_1$ be $\rho_{p1}$,
and let the star over-expand to radius $R_2$, while its luminosity $L$
stays constant. The surface temperature decreases
by a factor $T_{p2}=T_{p1}(R_2/R_1)^{-1/2}$, and
the new photospheric density is, by equation (6),
$\rho_{p2}=\rho_{p1}(R_2/R_1)^{-5.5}$.
The ratio of the density at $r=R_1$ in the over-expanded
to normal state is
\begin{equation}
\frac {\rho_2(r=R_1)}{\rho_{p1}}=
\frac{\rho_{p2}(R_1/R_2)^{-d}}{\rho_{p1}}
=\left(  \frac{R_2}{R_1} \right)^{d-5.5} .
\end{equation}

The last equation implies that if $d \lesssim 5.5$ then in the
outer parts of the envelope at $r \lesssim R_1$, the density
decreases as the star over-expands.
Inward to some radius $R_i$, mass shell contracts, releasing
gravitational energy.
The total envelope mass above $R_{\rm in}$ is
\begin{equation}
M(r>R_i)= \int _{R_i}^{R_p}
4 \pi \rho_p \left( \frac{r}{R_p} \right)^{-d} r^2 dr =
\frac {4 \pi}{d-3} R_p^d \rho_p (R_i^{3-d}-R_p^{3-d}).
\end{equation}
The radius $R_i$ is given by equating this mass between the
normal and over-expanded states.
Substituting for the photospheric radius $R_p$ and density $\rho_p$ for the
two states in the temperature range given in equation (6),
and using $\rho_{p2}=\rho_{p1}(R_2/R_1)^{-5.5}$, we find the radius $R_i$
\begin{equation}
R_i= \left(
\frac {1 - C_r^{-2.5}}{1 -C_r^{d-5.5}}
\right)^{\frac{1}{3-d}} R_1,
\end{equation}
where $C_r \equiv R_2/R_1$.

To estimate the value of $R_i$ I take $d=4.5$, as mentioned above.
As an example consider an over expansion from a photospheric
temperature of $2900 \K$ to $2200 \K$. The radius increases
by a factor of $C_r=R_2/R_1=1.7$, and for $d=4.5$ we find
$R_i= 0.68 R_1$.
Namely, because for these parameters the density in the outer part
of the envelope is lower in the over-expanded state (eq. 9),
in the regions close but inward to $R_i$ mass shells contract,
and release gravitational energy which may help
pushing the over-expanding layers above.
Exact solution of this process is required to determined the total
energy budget.

The shallow envelope structure, a necessary condition for
the proposed over-expansion process, is presented in Figure 1.
Presented are the temperature and radius as function of the
mass inward to the photosphere, i.e., the mass is zero at
the photosphere.
The AGB model is from Soker \& Harpaz (1999; see their figure 2;
for correct density scale see their figure 6).
The total envelope mass is $0.3 M_\odot$, approximately the stage when
over-expansion process might start.
With this figure and the previous discussion, the following should
be noted.
(1) For the over-expansion to occur, only the photosphere should
cool to $T \lesssim 3000 \K$. The regions inward to the photosphere
will be hotter, as in Figure 1.
(2) The shallow temperature and density profiles imply that
there are no dramatic changes in the structure of the outer envelope
as the envelope expands.
(3) Even with a mass loss rate as high as
$\dot M_{OE} \sim 10^{-4} M_\odot \yr^{-1}$, and an expansion
time as short as $\sim 100 \yr$ (see next section), the
total mass lost is $\sim 0.01 M_\odot$.
Within this mass layer the temperature is still very low,
as can be seen in figure 1. Hence, even the high mass loss rate
will not cause the envelope to go through dramatic changes.
The dramatic changes are in the mass loss rate and geometry,
resulting from the low effective (photospheric) temperature.
(4) In any case, the behavior of the inner parts of the envelope
may limit the over-expansion, such that the photosphere will not
cool all the way down to $T_p \sim 2000 \K$.
(5) During the very last stages of the AGB the contraction of the star
is significant. On one hand the now higher photospheric temperature
($T>3000 \K$) reduces the likelihood of the over-expansion process,
while on the other hand the envelope density profile is very shallow
(Soker \& Harpaz 2002), suggesting that gravitational energy
can be liberated by an over-expansion.
In Soker (2004) I suggested that such long-term oscillations,
between the over-expanded and normal states, can last until the
envelope mass reduces to $M_{\rm env} \simeq 0.02 M_\odot$
(depending on stellar luminosity and composition).
(6) To test the proposed process numerically, the stellar code
should include pulsations in the entire envelope, should
treat the convection and the photosphere.
It is very likely that the over-expansion will be triggered only
by pulsations.
Such numerical codes will be available in the near future;
presently most stellar codes include pulsations as an input, or calculate
pulsations but not the exact envelope structure and mass loss process.
Such numerical codes should also be able to treat the star for at least
$\sim 100~$years, and to start with a very low mass envelope which has
a very steep entropy gradient (Soker \& Harpaz 1999).
It is not clear that the numerical code of Freytag (2003; H\"ofner et al. 2005)
can handle these AGB stars with very little mass in their envelope.
However, when such codes get to be more accurate and capable for few
hundred years, I predict that they will find another semi-stable
configuration.

It should be emphasized that there is one stable (not
considering pulsations) structure for a considered AGB star.
I do not suggest that there is another stable structure.
What I suggest is that for low surface temperatures and as the low-mass
envelope there is a {\it local} minimum in the envelope energy, where
the envelope structure is somewhat more stable than for temperatures
slightly above or below this surface temperature.
I then speculate that under strong disturbances, such as
strong pulsations, the envelope can move to that position, stay
there for some time, $\sim 10-100$~yr, and then move back to its
globally stable structure.
The return to the stable position might be expedite by mass loss
(see next section). The ``tunneling'' of the envelope to this local
stability structure, if exists, requires strong disturbances,
hence it can be simulated only with numerical codes that include pulsations,
shocks, and molecule formation.
It is possible that only very strong perturbations, such as those
formed in the chaos mechanism proposed by Icke et al. (1992)
for irregular pulsating AGB stars, can cause this tunneling.

To qualitatively demonstrate the tunneling behavior, I build a very
simple toy model.
I consider the envelope to be split into three parts.
The very inner part contains most of the envelope mass, and
it does not responds to the change in the outer radius.
The outer part is averaged by gas of density $\rho_p$ residing
at radius $R_p$, with a total mass of
$\Delta M =4 \pi R_p^3 \rho_p/3$.
The intermediate part is average by radius $\zeta R_p$, where $\zeta<1$, and
mass of $M_f-\Delta M$, where $M_f$ is some small fraction of the envelope
mass.
The gravitational energy of the envelope as the envelope
expands or contracts is
\begin{equation}
E_{G{\rm env}} =
-\frac{ G M_\ast \Delta M}{R_p}
-\frac{ G M_\ast (M_f-\Delta M)}{\zeta R_p}
+C_{G1}
\end{equation}
where $C_{G1}$ is a constant.
Substituting for $\Delta M$, expressing $\rho_p$ as in
equation (2) with the same numerical values there, and expressing
$R_p$ as function of $T_p$ for $L=10^4 L_\odot$,
equation (l2) reads
\begin{equation}
E_{G{\rm env}} =
\frac{ G M_\ast M_f}{370 \zeta R_\odot}
\left[\frac{2.2\times 10^{-3} }{M_f/M_\odot}
\left( \frac{\kappa}{0.01 \cm^2 \g^{-1}} \right)^{-1}
\left( \frac{T_p}{3000 \K} \right)^{-3}(1-\zeta) -1 \right]
\left( \frac{T_p}{3000 \K} \right)^{2}
+C_{G1}
\end{equation}

The envelope energy of the toy model is plotted on figure 2,
in units of $G M_\ast M_f/370 \zeta R_\odot$, and for $\zeta=0.5$.
The constant $C_{G1}$ is arbitrarily set to
$C_{G1}=1.3G M_\ast M_f/370 \zeta R_\odot$.
Three lines are shown, for $M_f=0.06M_\odot$ (lower line),
$0.03M_\odot$, and $0.015M_\odot$ (upper line); recall that $M_f$ is only a
small fraction of the envelope mass in the toy model.
The three regimes of the opacity dependence on the temperature (eq. 5),
are seen by the broken discontinues line slope.
Since this is a toy model, the numerical values should not be
taken too seriously, only the qualitative behavior.
When the envelope mass is large the AGB stars has one stable solution.
This is true for low mass envelope as well.
However, in the low envelope mass case, a local minimum exists at much
lower photospheric temperatures $T_p$.
For example, when the stable structure for $M_f=0.01 M_\odot$ has
$T_p \sim 3100 \K$ (region A on figure 2), large disturbances might
cause the envelope to restructure itself to be in region B,
with $T_p \sim 2000-2400$.
The envelope will eventually go back to region A, its globally stable
structure.
When in region B, high mass loss rate is expected, such that most of
these stars are obscured in the optical.

In evolving from $T_p \simeq 3000 \K$ to $T_p \simeq 2000 \K$
the stellar radius increases by a factor of $\sim 2$.
The much lower temperature and larger radius will result
in a much higher mass loss rate (e.g., H\"ofner \& Dorfi 1997;
Wachter et al.\ 2002).
Using the expression derived by Wachter et al. (2002) I find the
mass loss rate to increase by more than an order of magnitude,
reaching a value of $\dot M_{OE} \sim 10^{-4} M_\odot \yr^{-1}$
at the maximum over-expanded state when $T_p \simeq 2000 \K$.
Inclusion of non-linear effects, e.g., magnetic cool spots
(see next section) or shocks produced by convective cells,
may further increase the mass loss rate.
Possible implications of over-expansion are the subject of
the next section.

\section{POSSIBLE IMPLICATIONS OF OVER-EXPANSION}

\subsection{Multiple semi-periodic concentric arcs (rings; shells)}

The arcs and rings which appear in the images of several PNs and
proto-PNs (Sahai {\it et al.} 1998; Kwok, Su, \& Hrivnak 1998;
Su {\it et al.} 1998; Sahai {\it et al.} 1999; Bond 2000;
Hrivnak, Kwok, \& Su 2001; Corradi et al. 2003; Corradi et al.\ 2004),
as well as in one AGB star (IRC +10216; Mauron \& Huggins 1999, 2000),
are thought to be concentric (more or less) semi-periodic shells;
some shells are complete while others are not.
Reviews of the arcs' and rings' properties are given by Hrivnak et al.,
(2001), Kwok et al.\ (2001), and Corradi et al.\ (2004).
To distinguish them from shells formed by other processes, e.g.,
thermal pulses, I term them multiple-arcs, or M-rings, although in
three dimensions they are shells, or fractions of shells.
The time intervals between consecutive arcs vary from system to system
and in some cases between arcs in the same system, with typical time
intervals of $t_s \sim 100-1000 \yrs$.
The rings and their spacing are expected to evolve with time
(Meijerink et al.\ 2003).
Theoretical models for their formation, e.g.,
instabilities in dust-gas coupling in the circumstellar matter
(Deguchi 1997; Simis, Icke, \& Dominik 2001; ),
a solar-like magnetic activity cycle in the progenitor AGB star
(Soker 2000a; Garc\'ia-Segura, Lopez, \& Franco 2001), as well as
other models, are discussed in detail by Soker (2002a). 
In that paper I concluded that models
that attribute the semi-periodic arcs to semi-periodic variation in
one or more stellar properties are most compatible with observations.
The magnetic activity cycle was the favorite such mechanism at the time.
The over-extended envelope may be connected to the formation
of the multiple-arcs.
Some multiple-arc objects are carbon-rich, e.g., IRC +10216
(Mauron \& Huggins 1999, 2000; Fong, Meixner, \& Shah 2003);
as noted, carbon-rich AGB stars have
high opacity, hence larger radii, similar to the over-expanded
envelope of cool oxygen-rich AGB stars.
Now, I also raise the possibility that a semi-periodic oscillation
between the ``normal state'' of the AGB stellar envelope
and the over-expanded state studied in the previous section,
for oxygen-rich AGB stars, may be another candidate mechanism for the
formation of the multiple semi-periodic arcs.
This mechanism, based on molecular opacity, is not directly applicable
to carbon-rich stars, but I do note that inclusion of dust opacity in
carbon-rich AGB stars may lead to a similar behavior.
This is beyond the scope of the present paper.

As mentioned in the previous section, the star can rapidly
over-expand by a factor of $\sim 2$ in radius, as the
photospheric temperature drops from $\sim 3000 \K$
to $\sim 2000 \K$.
This increases substantially the mass loss rate.
The question is, what determines the period of the oscillation,
if it exists?
First, it may be that a magnetic activity cycle exists
(Soker 2000a; Garc\'ia-Segura et al. 2001), which causes this
over-expansion at maximum activity.
Another possibility that may occur under specific conditions
is that the high mass loss rate from the over-expanded star
sets the periodicity.
 The typical mass in the maximum over-expanded part of the envelope,
when $T_p \simeq 2000 \K$,
is $M_{OE} \simeq 4 \pi \beta R_{OE}^3 \rho_p$, where $R_{OE}$
is the stellar radius in the over-expanded state, and $\beta \sim 1$.
Using the scaling in equation (6) gives
$ M_{OE} \simeq 0.02 M_\odot$.
For a mass loss rate $\dot M_{OE}$, this part of the envelope
will be depleted in a time
\begin{equation}
\tau_{OE} = \frac {M_{OE}}{\dot M_{OE}}
= 200 \frac{M_{OE}}{0.02 M_\odot}
\left( \frac{\dot M_{OE}}{10^{-4} M_\odot \yr^{-1}} \right)^{-1}
\yr .
\end{equation}
The mass loss rate of $\sim 10^{-4} M_\odot \yr^{-1}$ is observed
at the upper AGB, and it is in accord with Wachter
et al. (2002) for these parameters.
The depletion time of the over-expanded envelope may set the period
of the semi-periodic arcs: a large fraction of the over-extended
envelope mass is lost in the wind, such that the star shrinks and
mass loss rate drops.
When the photospheric temperature rises again to $\sim 3000 \K$,
another over-expanded phase starts.
The exact conditions for the occurrence of this speculative cycle should
be determined by stellar evolutionary numerical code, possibly
including rotation and magnetic activity.

\subsection{The transition to axisymmetric mass loss}

The much cooler over-expanded envelope is more susceptible
to some mechanisms which enhance dust formation, e.g., shocks
by convective cells 
and magnetic cool spots.
In particular, it is enough that during the over-expanded
state the temperature above magnetic cool spots be only slightly
lower to substantially enhance dust formation, hence mass loss rate.
If magnetic spots are concentrated in the equatorial plane,
this may increase the mass loss rate there, leading to axisymmetric
mass loss.
Because the total mass loss rate increases as well
(see previous subsection), this leads to a positive correlation
between mass loss rate and deviation from spherical mass loss.

Another effect is in systems where a companion at a large
orbital separation accretes from the AGB wind, forms an accretion
disk, and blows collimated fast wind (CFW; or two jets).
The higher mass loss rate and slower wind velocity, because
of lower escape velocity, will make such a mechanism much
more likely to occur (Soker 2001). 

\subsection{Binary AGB stars with circumbinary disks}
There is a group of post-AGB stars (or stars about to leave the AGB),
which have a binary companion and a circumbinary disk;
most of these stars are classified as RV Tauri stars,
(Mass et al.\ 2005; De Ruyter et al.\ 2005), and most systems have
their orbital period in the range $\sim 200-1500$~days (Mass 2004).
The orbital separation is $\sim 1-3 \AU$.
The circumbinary disk in most, or even all, of these systems are
oxygen rich.
The evolutionary puzzle is the formation mechanism of the disk.
It requires a strong interaction between the binary stars
to through material from the AGB envelope such that it has
enough angular momentum to form a disk, yet it does not
reach the escape velocity.
I propose that the disks are formed from mass loss episodes during
the periods when the AGB star experiences over-expanded states.
In these states the envelope reaches the companion, such that the
interaction of the companion with the envelope is strong enough to
form the circumbinary disk.
In some cases a short phase of common envelope is formed.

\subsection{The fate of Earth and other close companions}

The over-expanded envelope may swallow planets and lower mass
objects.
These will spin-up the envelope and may lead to
enhanced magnetic activity and axisymmetric mass loss
(e.g., Soker \& Harpaz 1999).
In some cases the entrance of a planet to the
envelope of its AGB parent star is marginal.
An extended envelope, even if for a relatively short time of
$\sim {\rm few} \times 10^3$ years, may increase tidal interaction
(friction with the envelope material
is negligible at those low densities), by an amount
enough to bring the planet into the envelope.
The question of the destiny of the Earth is such a marginal case
(Rybicki \& Denis 2001).
An over-expanded sun during its final AGB phase
may supply the tiny enhancement in tidal interaction strength
required to cause the Earth to spiral-in inside the
envelope, and evaporate.
The great uncertainties in the processes that determine the exact
outcome of the Earth-sun system (Rybicki \& Denis 2001) make any
present attempt to calculate the outcome meaningless.

\section{SUMMARY}

This paper has addressed a few open equations in the formation
of elliptical PNs, with the goal of pointing out that the behavior
of the opacity in oxygen-rich (similar to solar abundance)
upper AGB stars may lead these stars to substantially expand,
an {\it over-expansion}, when their photospheric
(effective) temperature drops to $T_p \sim 3000 \K$.
The reason is that the opacity in the photosphere of oxygen-rich AGB
stars sharply increases as temperature decreases from
$T_p \sim 2900 \K$ to $T_p \sim 2000 \K$
(Alexander \& Ferguson 1994; Ferguson et al.\ 2005; see eq. 5 here).
The much higher opacity implies a much lower photospheric
density (eq. 6).
When the effective temperature is $T_P \sim 3000 \K$, the
photospheric density is relatively high (eq. 2), while
the average envelope density of these upper AGB stars, which have
large radii and have lost most of their envelope, is very low (eq. 3).
Because the photospheric density must be much below the average
envelope density, eventually the star contracts and becomes a
post-AGB star.

The behavior of the opacity quoted above enables the star
to lower the ratio of the photospheric to average
density, thereby stabilizing its envelope structure
by lowering its photospheric temperature to
$T_p \lesssim 2900 \K$ down to $T_p \simeq 2000 \K$ (eq. 7).
This causes over-expansion up to $\sim 2$ times its radius at
$T_p \sim 3000 \K$.
Because the envelope mass is already low at the proposed
over-expansion phase, and the mass loss rate is expected to
increase by more than an order of magnitude (sec. 3.1), this phase
is relatively short; it lasts for $\sim {\rm few} \times 1000 \yr$,
and it does not occur in all oxygen-rich stars.
Therefore, these stars are relatively rare.
They are expected to be hidden behind dense circumstellar dust,
so I do not suppose that these stars can be observed
directly.

The eruptive star \astrobj{V838 Mon} may have gone through an evolutionary
phase resembling the proposed over-expanded phase.
The evolution of temperature and radius of \astrobj{V838 Mon} is presented in
Tylenda (2005).
There are three distinct regimes in the evolution of the radius:
($i$) A fast increase of radius with time, ending at time 120 days
(fig. 2 in Tylenda 2005);
($ii$) A very slow increase of radius with time, lasting from 120 to 240
days
after eruption;
($iii$) Fast decline of radius with time, starting after 240 days.
The second phase starts with a fast drop in photospheric temperature
from $\sim 3000 \K$ to $\sim 2400 \K$, and ends at the lowest temperature
of $\sim 1750 \K$. I attribute the slow increase in radius in the second
phase to the behavior of the opacity.
If the opacity hadn't been so high near $\sim 2000 \K$, the contraction
phase would have started earlier.

Some possible implications of the over-expanded phase were discussed.
(1) The over-expanded phase may be connected to the formation of
semi-periodic concentric arcs (rings; shells), observed in some
PNs and AGB stars, as discussed in section 3.1.
(2) The over-expanded AGB star may be more susceptible to some processes
that cause axisymmetric, rather than spherical, mass loss
geometry, e.g., magnetic cool spots (sec. 3.2).
With the higher mass loss rate, this may explain the
positive correlation between the mass loss rate and the
transition to axisymmetric mass loss geometry.
(3) An over-expanded AGB star is somewhat more likely to swallow low
mass companions (sec. 3.4), another process which may lead to
axisymmetical mass loss geometry.

Some PNs affected by the proposed process and its implications,
many whom are oxygen-rich, i.e., their oxygen abundance is
$\gtrsim 1.5$ times their carbon abundance,
were mentioned in the text (sec. 1).

This paper has not dealt with carbon-rich stars, for which
the opacity behaves differently.
The opacity starts to rise already at $T_p \simeq 4000 \K$,
implying much larger average radii for carbon-rich AGB stars than
for oxygen-rich AGB stars (Marigo 2002, 2003) .
Because the expansion of carbon-rich AGB stars occurs at higher
temperatures, hence early AGB phase, carbon stars spend longer
time as large AGB stars, much longer than the expected over-expanded
phase of oxygen-rich AGB stars.
I speculate that including dust opacity in the study of cool
carbon-rich stars may lead to some similarities to the suggested
behavior of over-expanded oxygen-rich stars.

\acknowledgements
This research was supported by the Israel Science Foundation.

\end{document}